\documentclass[]{aa} 
 
\usepackage{graphicx}
\usepackage{txfonts}
\usepackage{natbib}
\bibpunct{(}{)}{;}{a}{}{,}

\begin{document}

   \title{Evidence from high mass X-ray binaries that Galactic WR components of WR+O binaries end their life with a supernova explosion
         }

   \author{D. Vanbeveren\inst{1}, N. Mennekens\inst{1}, E.P.J. van den Heuvel\inst{2}
           \and J. Van Bever\inst{1}
          }

   \institute{Astronomy and Astrophysics Research Group, Vrije Universiteit Brussel, Pleinlaan 2, 1050 Brussels, Belgium\\
              \email{dvbevere@vub.be}
							\and Anton Pannekoek Institute for Astronomy, University of Amsterdam, PO Box 94249, 1090GE Amsterdam, the Netherlands
             }

   \date{Received 4 December 2019 / Accepted March 2020}

  \abstract
   {Theoretical population number studies of binaries with at least  one black hole (BH) component are obviously depending on whether or not BHs receive a (natal) kick during their formation.}{Several observational facts seem to indicate that indeed BHs receive a kick during their formation. In the present paper we discuss additional evidence.}{The progenitors of wind fed high mass X-ray binaries (HMXB) with a BH component (BH HMXB) are WR+OB binaries where the Wolf-Rayet (WR) star will finally collapse and form the BH. Starting from the observed population of WR+OB binaries in the Solar Neighborhood we predict the population of wind fed BH HMXBs as a function of the BH-natal kick.}{The simulations reveal that when WR stars collapse into a BH with zero or low kick, we would expect 100 or more wind fed BH HMXBs in the Solar Neighborhood whereas only one is observed (Cyg X-1). We consider this as evidence that either WR components in binaries end their life as a neutron star or that they collapse into BHs both accompanied by a supernova explosion imparting significant (natal) kicks.}{}

            
   \titlerunning{How do WR binary components end their life?}
   
   \authorrunning{Vanbeveren et al.}

   \maketitle

\section{Introduction}

The detection by LIGO/Virgo of gravitational waves (GW) resulting from merging binary neutron stars (BNS), merging binary black holes (BBH) and merging mixed binary systems (BNSBH) opened a new window to explore the Universe. One of the striking results (striking at least as far as the evolution of massive binaries is concerned) is that the LIGO/Virgo data support the existence of BHs with a mass up to (and even larger than) 40 M$_{\odot}$. Note that these high masses were predicted in the recent past (De Donder \& Vanbeveren, 2003, 2004; Belczynski et al., 2010) by evolutionary simulations of massive binaries where during core helium burning (= the Wolf-Rayet [WR] phase) the effect of stellar wind mass loss is calculated using a formalism obtained with a WR-atmosphere code that includes clumping and with a metallicity dependency as predicted by the radiatively driven wind theory (e.g., Hamann \& Koesterke, 1998, 2000; see also the Appendix).

The LIGO-discovery of the first massive BBH merger (GW150914) was announced in 2015. Note that before 2015 binary population number (BPN) simulations of double compact star binaries predicted GW150914-like events, e.g. Voss \& Tauris (2003), Dominik et al. (2012, 2013), Mennekens \& Vanbeveren (2014). Since 2015 such BPN simulations have been booming.

One of the main uncertainties in BPN prediction of BBH and BNSBH mergers is whether BH formation is accompanied by a supernova (SN) explosion and whether this SN explosion produces a BH kick that is large enough to affect overall BBH and BNSBH formation. This will be the main scope of the present paper.

The proper motion of pulsars has been intensively studied since the early nineties (e.g., Lyne \& Lorimer, 1994; Hansen \& Phinney, 1997; Arzoumanian et al., 2002; Pfahl et al., 2002; Hobbs et al., 2005) and this resulted in the now generally accepted conclusion that neutron stars (NSs) may receive kicks at birth (natal kicks - NKs) with an average in the range 200-500 km/s, when they are formed in core-collapse SNe. Possible exceptions are NSs born in a prompt (fast) electron-capture SN where the resulting kick is expected to be small (Nomoto, 1984, 1987; Podsiadlowski et al., 2004).

Arguments favoring or disfavoring BH kicks based on direct physical principles are inconclusive (Ozel et al., 2010; Farr et al., 2011; Belczynski et al., 2012). Individual binaries containing a BH companion have been investigated by many authors (e.g., Brandt et al., 1995; Nelemans et al., 1999; Remillard et al., 2000;  Gualandris et al., 2005; Willems et al., 2005; Dhawan et al., 2007; Fragos et al., 2009; Wong et al., 2010, 2012, 2014). A common conclusion is that the BHs in these binaries must have been formed without or with rather modest NKs.

Repetto et al. (2012) (see also Repetto et al., 2017) considered the observed distribution of distances above the Galactic plane of low-mass X-ray binaries containing a BH component and compared this with binary population synthesis simulations by adopting different initial NK-distributions. They conclude that a distribution similar to that of neutron stars (the one of Hanson \& Phinney, 1997) seems to be preferred. For a critical discussion of the Repetto et al. results see Belczynski et al. (2016).

Woosley (2019) presented detailed evolutionary computations of massive galactic helium stars that have lost all their hydrogen at the beginning of core helium burning (e.g. remnants after Roche lobe overflow (RLOF) if these stars were binary components = He-zero age main sequence (He-ZAMS) stars).  To account for mass loss by Wolf-Rayet-like stellar winds the author used the prescription of Yoon (2017) (see Appendix). An important conclusion resulting from these computations is that most of the hydrogen free helium stars with He-ZAMS mass between 9 M$_{\odot}$ and 60 M$_{\odot}$ (corresponding to end-core-helium burning mass between 7 M$_{\odot}$ and 30 M$_{\odot}$ and to progenitor H-ZAMS mass between 30 M$_{\odot}$ and 120 M$_{\odot}$) make BHs. It is unclear whether some of them might explode and for those that explode whether the explosion is symmetric or asymmetric.

To further study the latter the following may be a possibility. The 9-60 M$_{\odot}$ He-ZAMS mass range is also the mass range of the WR stars in the known galactic WR+O binaries. If the WR star collapses into a BH, a BH+O binary is formed which may become a wind fed high mass X-ray binary (a BH HMXB like Cyg X-1). In the present paper we will consider the following question: there are about 70 WR+OB binaries and probable WR+OB binaries in the Solar Neighborhood\footnote{3-4 kpc from the Sun is defined here as the Solar Neighborhood.} (van der Hucht, 2001) and (only) 1 high mass BH+OB X-ray binary. In how far is this compatible with the assumption that the WR stars collapse into a BH and that the collapse does not disrupt the binary? Section 2 describes the methodology. In sections 3 and 4 we demonstrate that the answer to the foregoing question is negative. Section 5 then deals with an alternative.

\section{Method}

We consider the population of observed WR+O binaries in the Solar Neighborhood and select those where we have a reasonable estimate of the masses of both components. Using our evolutionary computations of hydrogen deficient post-RLOF massive core helium burning stars\footnote{We use the Brussels stellar evolutionary code as it has been described in Vanbeveren et al. (1998c). One of the main uncertainties affecting significantly the core helium burning evolution of massive stars and binary components, is the stellar wind mass loss rate formalism used in the code. The latter is discussed in the Appendix.}, given the mass of the WR star, we calculate the expected mass of the WR star at the end of core helium burning. To do this we assume that in case the WR star is of the nitrogen sequence (a WN star) (resp. of the carbon sequence; a WC star) the star is at the beginning of the core helium burning phase (resp. beginning of the WC phase defined as the moment during core helium burning where due to stellar wind mass loss 3-alpha burning products appear at the surface). Remark that this assumption does not significantly affect the overall conclusions of the present paper. If the final WR star collapses into a BH without a SN explosion (i.e. the mass of the BH = the final WR mass) a BH+O binary is formed that may evolve into a (wind fed) BH HMXB. Given the mass-spectral type-luminosity class of the O-type companion we simulate its further evolution by interpolating in the Geneva tracks described in Ekstrom et al. (2012). Note that these evolutionary computations hold for massive stars that rotate at rates which are similar as the rotation rates of O-type components in WR+O binaries (Shara et al., 2017). We now proceed as follows. A massive binary system consisting of an OB star and a BH may become a HMXB when part of the stellar wind matter lost by the OB star is trapped gravitationally by the BH. Losses of gravitational energy of this material at the time of accretion by the BH then produce the X-ray radiation.

To obtain X-rays, however, the presence of an accretion disk is required in case of a BH (Shapiro \& Lightman, 1976; Iben et al., 1995). A Keplerian disk is formed when the specific angular momentum of the accreted matter exceeds that of the largest possible stable orbit around the BH (corresponding to a radius $R_K$, with $R_K$ approx. 3 times the Schwarzschild radius). This condition is satisfied when

\begin{equation}
\frac{R}{A} \geq \left(1-\frac{R}{A}\right)^{\frac{8}{7}}\left(\frac{R_K}{R}\right)^{\frac{1}{7}}\left(\frac{M}{M_{BH}}\right)^{\frac{3}{7}}
\end{equation}

\noindent or equivalently

\begin{equation}
\frac{R}{A} \geq \frac{a}{1+a}
\end{equation}

\noindent with $a=\left(\frac{R_K}{A}\right)^{\frac{1}{8}}\left(\frac{M}{M_{BH}}\right)^{\frac{3}{8}}$, with $R$ and $M$ being the radius resp. the mass of the optical component, $A$ is the semi major axis of the binary and $M_{BH}$ the mass of the BH. When the optical OB companion of a BH is not filling its Roche lobe, the latter accretes material by capturing a fraction of the stellar wind gas of the former. To estimate the X-ray luminosity in these wind-fed systems, we use the model of Davidson \& Ostriker (1973), involving accretion of matter according to Bondi \& Hoyle (1944).

We identify the X-ray luminosity $L_\mathrm{X}$ with the gravitational energy release of the accreted matter

\begin{equation}
L_\mathrm{X} = G\frac{M_{BH}\dot{M}_{accr}}{R_K}
\end{equation}

\noindent where $G$ is the gravitational constant and $\dot{M}_{accr}$ is the amount of matter accreted from the stellar wind by the BH in a unit of time, given by

\begin{equation}
\dot{M}_{accr}= \varepsilon \left(\frac{M_{BH}}{M}\right)^2\frac{\delta^2}{(1+\delta)^{3/2}}\dot{M}
\end{equation}

\noindent with $\delta=\frac{R}{2A}\left(\frac{v_{esc}}{v_{wind}}\right)^2$. Here, $v_{esc}$ and $v_{wind}$ are the escape velocity of the OB-star resp. wind velocity at the location of the BH. The stellar wind mass loss rate of the OB-type star, $\dot{M}$ is calculated using the formalism proposed by Vink et al. (2000), corresponding with the mass loss rate formalism used in the Geneva evolutionary code.

As shown by Groenewegen \& Lamers (1989) $v_{wind}$ can be described by a $\beta$-law $v_{wind}=v_\infty\left(1-\frac{R}{A}\right)^{\beta}$ with $v_\infty$ the wind velocity at infinity and $\beta$ between 0.7 and 1.5. Our results are computed with $\beta = 1$ but the $L_\mathrm{X}$ values differ by no more than a factor of 2 when other values of $\beta$ are used and this difference is small enough not to affect our basic conclusions. The ratio $\frac{v_{esc}}{v_\infty}$ is taken from the formalism also proposed by Vink et al. (2000).

The factor $\varepsilon$ in equation (4) accounts for a self-limiting effect on the X-ray luminosity, due to its radiation pressure. Eddington (1926) estimates it as

\begin{equation}
\varepsilon = \left(1-\frac{L_\mathrm{X}}{L_{Edd}}\right)^2
\end{equation}

\noindent with $L_{Edd} = \frac{4\pi cGM_{BH}}{\sigma_e}=\frac{65335L_{\odot}}{1+X}\frac{M_{BH}}{M_{\odot}}$, where $X$ represents the hydrogen abundance in the accreted material, $c$ is the speed of light, and $\sigma_e$ the electron scattering coefficient. The mass accretion rate $\dot{M}_{Edd}$ producing $L_{Edd}$ through its gravitational energy release satisfies

\begin{equation}
L_{Edd} = G\frac{M_{BH}}{R_K}\dot{M}_{Edd}.
\end{equation}

\noindent Then, solving for $L_\mathrm{X}$ one obtains

\begin{equation}
L_\mathrm{X} = \frac{\alpha}{\left(1+\sqrt{1+\alpha}\right)^2}L_{Edd}
\end{equation}

\noindent with $\alpha = 4 \frac{\delta^2}{(1+\delta)^{3/2}}\left(\frac{M_{BH}}{M}\right)^2\frac{\dot{M}}{\dot{M}_{Edd}}$.

To illustrate our method let us consider a fictitious but typical WR+O binary: a WN+O6V system with masses 19 + 32 M$_{\odot}$ and with an orbital period of 2, 4, 10 and 15 days. The pre-RLOF progenitor of the WN star has a ZAMS mass of 40 M$_{\odot}$ implying that the WN star is $\sim$5.5 Myr old (see Vanbeveren et al., 1998b, c). At the end of core helium burning ($\sim$0.5 Myr later) the WR star is a WC type star with mass = 11 M$_{\odot}$ whereas the orbital period has increased due to the spherical wind and reaches values of resp. 2.8, 5.6, 14.1 and 21.1 days. According to the spectral type and luminosity class of the O-type companion the star is on the Geneva 32 M$_{\odot}$ evolutionary track somewhere between 0 and 2 Myr (i.e. due to mass accretion during the previous RLOF the O star is rejuvenated and looks like a 32 M$_{\odot}$ star that is at most 2 Myr old although the lifetime of the binary is 5.5 Myr). We begin our simulations starting from the 2 Myr point on the 32 M$_{\odot}$ evolutionary track and we compute the temporal evolution of the X-ray luminosity using the formalism described above and assuming that the 11 M$_{\odot}$ WC star collapses into a 11 M$_{\odot}$ BH without SN explosion and thus without a natal kick. Results are shown in figure 1\footnote{The final rise of the X-ray luminosity is due to the fact that the 32 M$_{\odot}$ star enters the hydrogen shell burning phase implying a rapid increase of the radius and of the stellar wind mass loss rate (the Vink-rates) of the star. Due to the fact that this phase is very short it does not affect our population number results. The simulations stop when the star fills its Roche lobe.}. First note that all our wind fed systems have $L_\mathrm{X} > 10^{35}$ erg/sec which means that they would have been detected by present (and previous) all-sky-monitors if the sources are located within 3-4 kpc from the Sun (prof. Wijnands, prof. van der Klis, private communications. Note that High-Mass X-ray Binaries with X-ray luminosities similar to the ones of the BH-systems predicted here (around $10^{35}$ erg/s  or larger) are concentrated close to the Galactic plane, and are being seen throughout our Galaxy. So, extinction is not expected to play a role in their detectability. Particularly the hard part of their X-ray spectrum (above 10 keV), which tends to dominate in the case of BHs, is not expected to suffer any interstellar extinction).  The results can be interpreted as follows: the average core helium burning lifetime of the WR+O binary is 400000 $\pm$ 100000 yrs. Figure 1a (where we see that the X-ray lifetime of the wind fed system is 4-5 Myr) then illustrates that for every WR+O binary within 3-4 kpc from the Sun with masses 19+32 M$_{\odot}$ and with a period of 2 days we expect to observe 4-5 Myr/300000-400000 yr = 10-17 BH+OB HMXBs. When we repeat this for the same WR+O binary but with an orbital period of 10 days, it follows from figure 1d that we expect at most 1 BH+OB HMXB (note that for a WR+O period of 15 days we essentially expect no wind fed HMXB). By considering the observed WR+O binary population within 3-4 kpc from the Sun, and performing the previous exercise for all of them, we can get an indication how many BH+OB HMXBs are expected within 3-4 kpc if it is assumed that the WR stars in all the WR+O binaries collapse into BHs without SN explosion. This will be discussed in the next section.

\begin{figure*}[]
\centering
   \includegraphics[width=12.6cm]{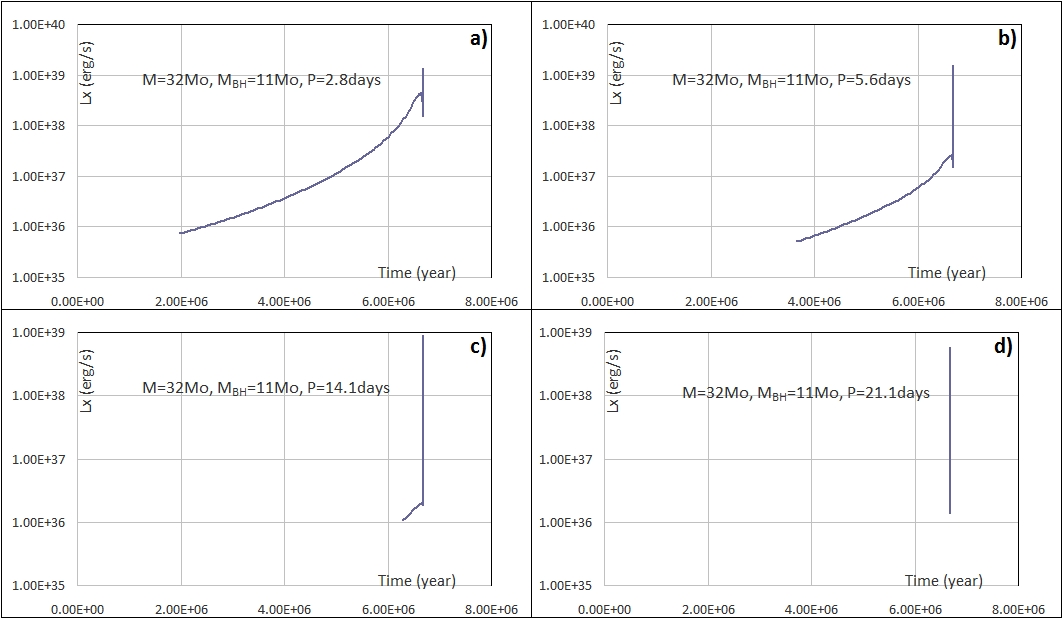}
     \caption{The temporal evolution of the X-ray luminosity (as a function of O-star apparant, rejuvenated age; see also text) of an O+BH binary with orbital parameters indicated on the figure.}
     \label{fig:1}
\end{figure*}

\section{The WR+O binary population in the Solar Neighborhood}

Using the WR catalogues of van der Hucht (2001, 2006) and of Crowther (2017) Table 1 collects the double-lined spectroscopic WR binaries in the Solar Neighborhood for which reasonable estimates of the minimum masses of both components are known. The procedure as outlined by Vanbeveren et al. (2018) allows to propose most probable masses of both components. Summarizing: applying the mass-spectral type/luminosity class relation proposed by Vanbeveren et al. (1998c) for the O-type component yields the mass-range estimates of Table 1\footnote{In case the luminosity class is not known assuming class V applies for the minimum mass, assuming class I applies for the maximum mass.}. Note that very similar results are obtained when the calibration of stellar parameters of Galactic O stars of Martins et al. (2005) is used. The combination with the observed mass functions and the inclination angle ranges (when available we used the values given in the catalogues listed above) explains the adopted masses in Table 1. Using the method outlined at the beginning of the previous section then allows to propose the masses of the WR components and the system periods at the end of core helium burning. Two remarks are appropriate:
\begin{enumerate}
\item The minimum mass of the WR components $\sim$9 M$_{\odot}$ (corresponding to an initial hydrogen ZAMS mass $\sim$25-30 M$_{\odot}$)
\item Accounting for the study of Woosley (2019) it cannot be excluded that most of the WR components will end their life as a BH.
\end{enumerate}

\section{The expected BH+O wind fed HMXB population in the Solar Neighborhood}

We calculate the expected BH+O wind fed HMXB population in the Solar Neighborhood starting from the WR+O sample of Table 1 and assuming that at the end of core helium burning the WR star collapses into a BH without SN explosion. The further evolution of the O component (and thus the further evolution of the BH+O binary) is interpolated from the Geneva tracks. The onset of X-ray radiation and the evolution of the X-ray luminosity is then computed using the method of section 2. Table 1 lists the lifetime where the system is visible as a wind fed HMXB with $L_\mathrm{X} > 10^{35}$ erg/s. As outlined in section 2, division by the average WR lifetime = 400000 yrs gives an indication how many wind fed BH+O HMXBs are expected for the WR+O binary considered. The table illustrates the conclusion that if the WR components in the 17 systems collapse into a BH without SN explosion, we expect 44 wind fed BH+O HMXBs in the Solar Neighborhood. If the 17 systems are representative for the whole population of WR binaries in the Solar Neighborhood ($\sim$80 according to the WR catalogues of van der Hucht, 2001, 2006 and of Crowther, 2017) the number of expected wind fed BH+O HMXBs becomes even more than 200. Since only one system is observed (Cyg X-1) the discrepancy is enormous.

\section{Two possible solutions}

\subsection{Most WR stars end their life as a NS}

Binary interaction (RLOF or the common envelope process) removes the hydrogen rich envelope of the mass loser on a very short timescale: the thermal timescale in case of a RLOF, the dynamical timescale in case of the common envelope process. At the end of RLOF most of the mass losers are hydrogen depleted stars at the beginning of core helium burning, i.e. in case of massive binaries the hydrogen depleted WR phase always starts at the beginning of core helium burning. Together with the effects of WR-like stellar wind mass loss this has a significant effect on the evolution of the convective helium burning core in general, on the evolution of the compactness of the core (the compactness parameter as introduced by O'Connor \& Ott, 2011) in particular. Detailed calculations of the variation of the compactness parameter in massive helium burning stars have been presented by Sukhbold et al. (2016) and Ebinger et al. (2019) (see also references therein). Although these computations hold for massive single stars, we think that they illustrate that it cannot be excluded that many of the WR stars in our Galactic WR+O binaries end their life as a NS. Since these NSs are expected to be accompanied by a asymmetric SN explosion, many of the WR+O binaries may be disrupted and it is clear that this would (at least partly) solve our problem.

\subsection{During their formation BHs receive a kick}

The large number of expected wind fed BH+O HMXBs discussed in the previous section obviously depends critically on the assumption that the collapse of the WR stars into a BH is not accompanied by a SN explosion. The discrepancy with observations may therefore be indicative that BH formation is accompanied by a SN explosion and that the BH receives a (natal) kick. If the natal kick is large enough to disrupt the binary, obviously no wind fed HMXB is expected. However, the binary does not need to be disrupted in order to reduce drastically the number of expected wind fed HMXBs. The simulations shown in figure 1 (section 2) illustrate that very few wind fed HMXBs are expected from BH+O binaries with a period $> 15$ days (see also the evolutionary models of the WR+O binaries in Tables 1). So, a natal kick velocity distribution that assures that most of the BH+O binaries resulting from the WR+O sample have a period larger than 15 days will significantly reduce the expected number of wind fed HMXBs, and explain the fact that only one (Cyg X-1) is observed. To get an idea of the order of magnitude of the kick velocities that are capable to do this we use the formularium given below.

We define

\begin{equation}
m = \sqrt{\frac{M_1+M_2}{M_{BH}+M_2}},
\end{equation}
\begin{equation}
v_{rel} = \frac{||\vec{v}_{kick}||}{v_{orb}},
\end{equation}
\begin{equation}
v_{orb} = \sqrt{G\frac{M_1+M_2}{A}}
\end{equation}

\noindent where $M_1$ and $M_2$ are the masses of both components before the collapse and $A$ is the distance between both components before the collapse. $v_{rel}$ thus represents the magnitude of the kick velocity relative to the orbital velocity. If $P$ is the binary period before the SN explosion, $P_{PS}$ the binary period after the SN explosion, $\theta$ and $\phi$ the angles that describe the direction of the kick velocity $\vec{v}_{kick}$ with respect to the orbital velocity $v_{orb}$ before the collapse + SN explosion of $M_1$, then following Brandt \& Podsiadlowski (1995) (see also Vanbeveren et al., 1998c)

\begin{equation}
c = \left(\frac{P\sqrt{m}}{P_{PS}}\right)^{\frac{2}{3}}=2-m(1+2v_{rel}\cos\theta \cos\phi+v_{rel}^2)
\end{equation}

\noindent and after some straightforward calculus

\begin{equation}
v_{rel} = -\cos\theta \cos\phi+\sqrt{-1+(\cos\theta \cos\phi)^2+\frac{2-c}{m}}.
\end{equation}

For the WR+O binaries listed in Table 1 we calculate the $v_{kick}$ value needed to obtain a BH+O binary with a period of 15 days (we also give the value to disrupt the binary). For most of the WR+O binaries $ m$ is close to 1 at the end of the core helium burning phase of the WR component. We intend to give indicative results and therefore we simplified the previous formulae by replacing $m$ by 1. Note that in this case an asymmetrical SN explosion (and a resulting substantial kick) is still possible due to the neutrino momentum flux (see e.g. Brandt \& Podsiadlowski 1995). Results for this case $m=1$ are given in Table 2 for $\cos\theta \cos\phi = 1$ (corresponding to a minimum kick, $v_{kick,min}$, that is needed to obtain the desired result) and $\cos\theta \cos\phi = 0$ (since $\cos\theta \cos\phi$ ranges between -1 and +1, the $v_{kick}$ corresponding to $\cos\theta \cos\phi = 0$ can be considered as an average value, $<v_{kick}>$). Similarly, a maximum value $v_{kick,max}$ can be calculated for $\cos\theta \cos\phi = -1$. The $<v_{kick}>$ values of Table 2 allow to conclude that the paucity of observed BH+O wind fed HMXBs in the Solar Neighborhood can at least partly be explained when the collapse of WR stars into a BH is accompanied by an asymmetric SN explosion resulting in a moderate kick imparted to the BH. We also notice that the $<v_{kick}>$ values that are needed to disrupt most of the WR+O binaries during the collapse of the WR star into a BH, are very similar to the average value of the kick velocity distribution of NSs. This means that when BHs are formed with kicks similar to those of NS (as suggested by Repetto et al., 2012, 2017) then most of the WR+O binaries will be disrupted when the WR star collapses into a BH.

\begin{table*}
\centering
\caption{For the selected observed WR+O binaries the mass-range of the O star based on its spectral type and luminosity class, the adopted WR and O masses based on all available data, the estimated mass of the WR star at the end of core helium burning, the present observed period and the estimated period at the end of the core helium burning phase of the WR star, the time (in Myr) where the resulting wind fed BH+O will be observable as a X-ray binary and finally the number of expected wind fed BH high mass X-ray binaries. Masses in M$_{\odot}$, periods in days.}
\begin{tabular}{c c c c c c c c c c c}
\hline
 & & O-star & adopted & adopted & mass at & & period & & \\
 & spectral & mass & mass & mass & end CHeB & present & at end & $L_\mathrm{X}$ time & $L_\mathrm{X}$ time\\
system & type & est. & (WR) & (O) & (WR) & period & CHeB & (Myr) & (Myr)/0.4 \\
\hline
WR21 & WN5+O4-5 & 37-60 & $>$19 & $>$37 & $>$10 & 8.3 & 11.8 & 0.3 & 0.8\\
WR30 & WC6+O6-8 & 24-50 & 16 & 34 & 14 & 18.8 & 20.4 & 0.0 & 0.0\\
WR31 & WN4+O8V & 24-34 & $>$11 & $>$24 & $>$7  & 4.8 & 6.1 & 2.5 & 6.3\\
WR35a & WN6+O8.5V & 19-33 & 18 & 19 & 10 & 41.9 & 68.2 & 0.0 & 0.0\\
WR42 & WC7+O7V & 27-37 & 16 & 27 & 14 & 7.9 & 8.7 & 1.8 & 4.5\\
WR47 & WN6+O5V & 37-70 & $>$40 & $>$47 & $>$20 & 6.2 & 10.5 & 0.6 & 1.5\\
WR79 & WC7+O5-8 & 24-60 & $>$10 & $>$24 & $>$7 & 8.9 & 10.7 & 1.1 & 2.8\\
WR97 & WN5+O7 & $>$30 & $>$17 & $>$30 & $>$9 & 12.6 & 18.3 & 0.0 & 0.0\\
WR113 & WC8+O8-9IV & 20-30 & $>$11 & $>$22 & $>$8 & 29.7 & 35.9 & 0.0 & 0.0\\
WR11 & WC8+O7.5III-V & 25-47 & 10 & 31 & 8 & 78.5 & 86.8 & 0.0 & 0.0\\
WR127 & WN3+O8.5V & 17-24 & $>$9 & $>$20 & $>$6 & 9.5 & 11.8 & 0.8 & 2.0\\
WR139 & WN5+O6III-V & 28-59 & 9 & 28 & 6 & 4.2 & 5.0 & 3.0 & 7.5\\
WR151 & WN4+O5V & 28-56 & 20 & 28 & 10 & 2.1 & 3.4 & 5.7 & 14.3\\
WR133 & WN5+O9I & 34-66 & 17 & 34 & 9 & 112.4 & 158.1 & 0.0 & 0.0\\
WR141 & WN5+O5III-V & 26-59 & 36 & 26 & 18 & 21.7 & 43.1 & 0.0 & 0.0\\
WR155 & WN6+O9II-Ib & 30-44 & 24 & 30 & 12 & 1.6 & 2.6 & 1.5 & 3.8\\
WR9 & WC5+O7 & $>$30 & 9 & 32 & 8 & 14.3 & 15.0 & 0.1 & 0.3\\
\end{tabular}
\label{tab:O-masses}
\end{table*}

\begin{table*}
\centering
\caption{For the WR+O binaries listed in Table 1 we calculate the $v_{kick}$ needed to obtain a BH+O binary with a period of 15 days (we also give the value to disrupt the binary). The meaning of $v_{kick,min}$, $<v_{kick}>$ and $v_{kick,max}$ is explained in the text. Velocities in km/s.}
\begin{tabular}{c c c c c c c c}
\hline
 & spectral & $v_{kick,min}$ & $v_{kick,min}$ & $<v_{kick}>$ & $<v_{kick}>$ & $v_{kick,max}$ & $v_{kick,max}$\\
system & type & $P=15$d & disrupted & $P=15$d & disrupted & $P=15$d & disrupted\\
\hline						
WR21 & WN5+O4-5 & 24 & 140 & 130 & 338 & 701 & 817\\
WR30 & WC6+O6-8 & - & 117 & - & 284 & - & 685\\
WR31 & WN4+O8V & 75 & 152 & 246 & 367 & 809 & 886\\
WR35a & WN6+O8.5V & - & 66 & - & 160 & - & 387\\
WR42 & WC7+O7V & 51 & 148 & 197 & 358 & 767 & 864\\
WR47 & WN6+O5V & 95 & 196 & 315 & 472 & 1039 & 1140\\
WR79 & WC7+O5-8 & 45 & 134 & 175 & 324 & 692 & 781\\
WR97 & WN5+O7 & - & 114 & - & 275 & - & 664\\
WR113 & WC8+O8-9IV & - & 83 & - & 201 & - & 485\\
WR11 & WC8+O7.5III-V & - & 68 & - & 164 & - & 395\\
WR127 & WN3+O8.5V & 20 & 115 & 107 & 278 & 576 & 671\\
WR139 & WN5+O6III-V & 94 & 167 & 292 & 405 & 903 & 977\\
WR151 & WN4+O5V & 132 & 198 & 378 & 478 & 1087 & 1153\\
WR133 & WN5+O9I & - & 57 & - & 138 & - & 334\\
WR141 & WN5+O5III-V & - & 89 & - & 215 & - & 519\\
WR155 & WN6+O9II-Ib & 162 & 224 & 448 & 540 & 1242 & 1303\\
WR9 & WC5+O7 & - & 123 & - & 296 & - & 715\\
\end{tabular}
\label{tab:WRkicks}
\end{table*}

\section{Final remarks}

\begin{enumerate}
\item Both solutions discussed in the previous section would explain in a straightforward manner the fact that the Solar Neighborhood only contains 1 WR+BH binary (Cyg X-3).
\item The number of merging BBH expected from isolated binary evolution would obviously be significantly reduced.
\item A kick may lead to an eccentric system and one may wonder if one could get a periodically wind fed system even when the binary period is 15 days or larger. However investigating the effects of eccentricity and of circularization is beyond the scope of the present paper since it will not affect the main conclusion, i.e., the population of high mass X-ray binaries in general, those with a BH component in particular seem to indicate that many Galactic WR stars in WR+O binaries end their life with an asymmetrical supernova explosion.
\end{enumerate}

\begin{acknowledgements}
      We thank the anonymous referee for valuable comments and suggestions.
\end{acknowledgements}

\appendix
\section{The mass loss rate formalism during the core helium burning phase of hydrogen depleted massive stars}

\begin{figure*}[]
\centering
   \includegraphics[width=16.8cm]{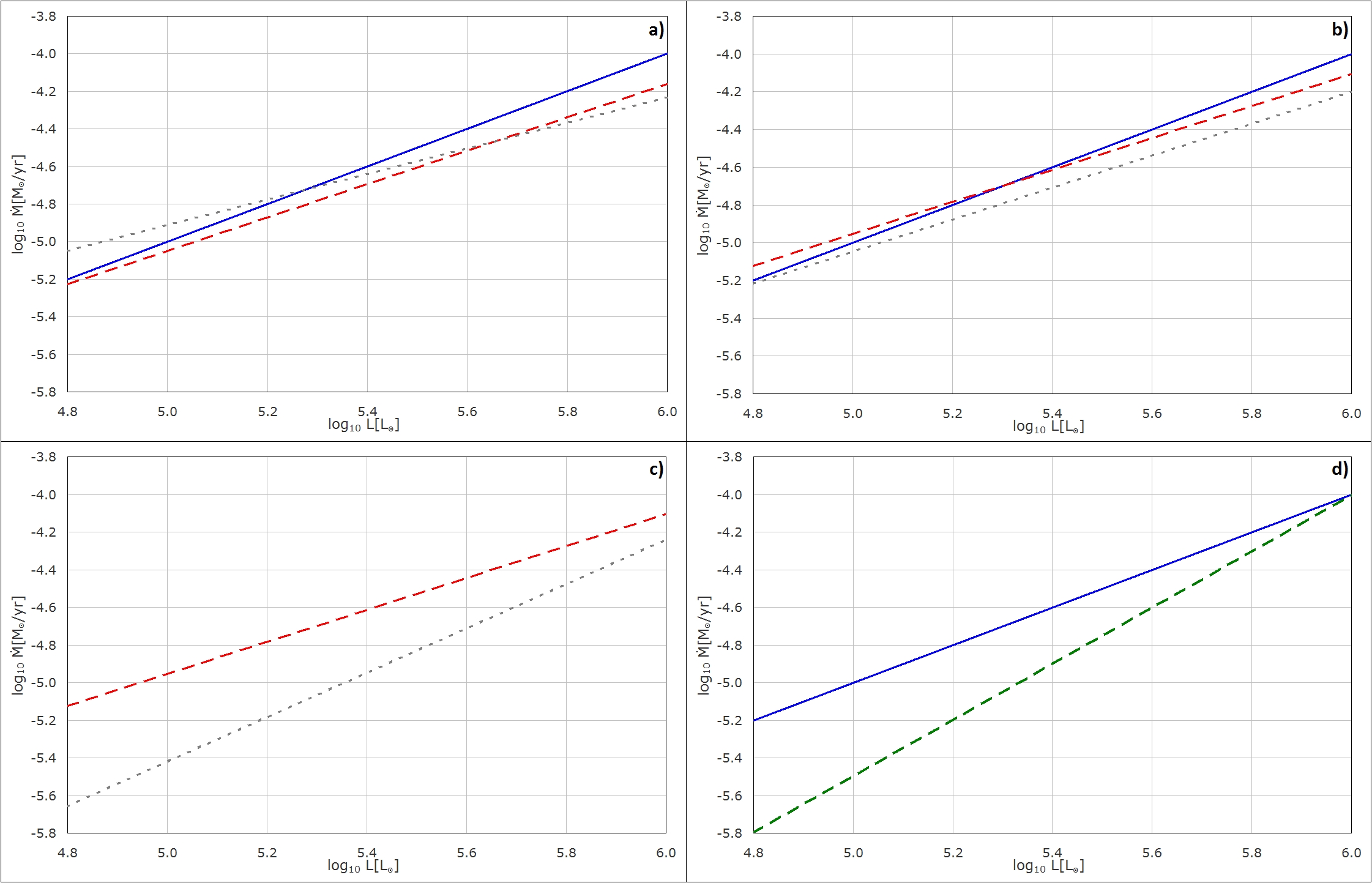}
     \caption{The stellar wind mass loss rate as function of luminosity of hydrogen depleted massive stars. \textbf{(a)} Solid blue: Brussels. Dashed red: WNE Potsdam. Dotted gray: WC Potsdam. \textbf{(b)} Solid blue: Brussels. Dashed red: Tramper WC $Y=0.98$. Dotted gray: Tramper WC $Y=0.6$. \textbf{(c)} Dashed red: Potsdam. Dotted gray: Yoon $f_{WR}=1$. \textbf{(d)} Solid blue: Brussels. Dashed green: Warsaw.}
     \label{fig:A}
\end{figure*}

To calculate the evolution of hydrogen depleted massive stars (singles and binaries) we (the Brussels team) use since 1998 (Vanbeveren et al., 1998a, b, c) a mass loss rate formalism during core helium burning that is based on the clumping corrected mass loss rates of WR stars that were available at that time (Hamann \& Koesterke, 1998) but also on the WN/WC number ratio of WR-binaries (a mass loss rate that is too large predicts a WN/WC number ratio that is too small and vice versa). We proposed

\begin{equation}
\log_{10}\dot{M} [\mathrm{M}_{\odot}/\mathrm{yr}] = \log_{10}L [\mathrm{L}_{\odot}] - 10.
\end{equation}

Recently, the mass loss rates of Galactic WN and of Galactic WC stars and WR stars of the oxygen sequence (WO stars) were reconsidered accounting for the impact of revised distances from Gaia DR2 (Hamann et al., 2019; Sander et al., 2019). In Figure A.1(a) we compare our linear relation with the linear regression lines of Hamann et al. and of Sander et al. The figure illustrates the statement that the core helium burning evolutionary results predicted with our mass loss rate formalism should be more than satisfactory in general, for the scope of the present paper in particular. The latter is strengthened by Figure A.1(b) where we compare our formalism with the formalism for WC and WO stars proposed by Tramper et al. (2016). And finally Figure A.1(c) compares the Potsdam formalism for early type WN stars (WNE stars) with the one advocated by Yoon (2017, $f_{WR} = 1$). As can be noticed the latter may significantly underestimate the true WNE-rates.

The relation between the initial mass of a massive primary in an interacting binary and its mass at the end of core helium burning (just prior to the formation of a NS or a BH) critically depends on the adopted stellar wind mass loss rate formalism during core helium burning. The relation computed with our mass loss rate formalism was presented for the first time by De Donder \& Vanbeveren (2003, 2004) and is shown once more in Figure A.2. The figure illustrates that when the mass of the BH in Cyg X-1 is confirmed to be 15 M$_{\odot}$, then the initial mass of the BH-progenitor should be at least 60 M$_{\odot}$. Moreover, massive BHs (like those predicted by the LIGO/Virgo data) are no surprise in low metallicity regions. We like to end this appendix with a word of caution. Number synthesis simulations that aim to predict the populations of double compact star binaries (NS+NS, NS+BH, BH+BH) rely on the evolution during core helium burning and thus on the WR-type stellar wind mass loss rate formalism. In Figure A.1(d) we compare our formalism with the one used by the Warsaw team (Belczynski et al., 2010). The difference is quite significant and may explain (at least partly) why some of the results of Warsaw differ from those of Brussels.

\begin{figure}[]
\centering
   \includegraphics[width=8.4cm]{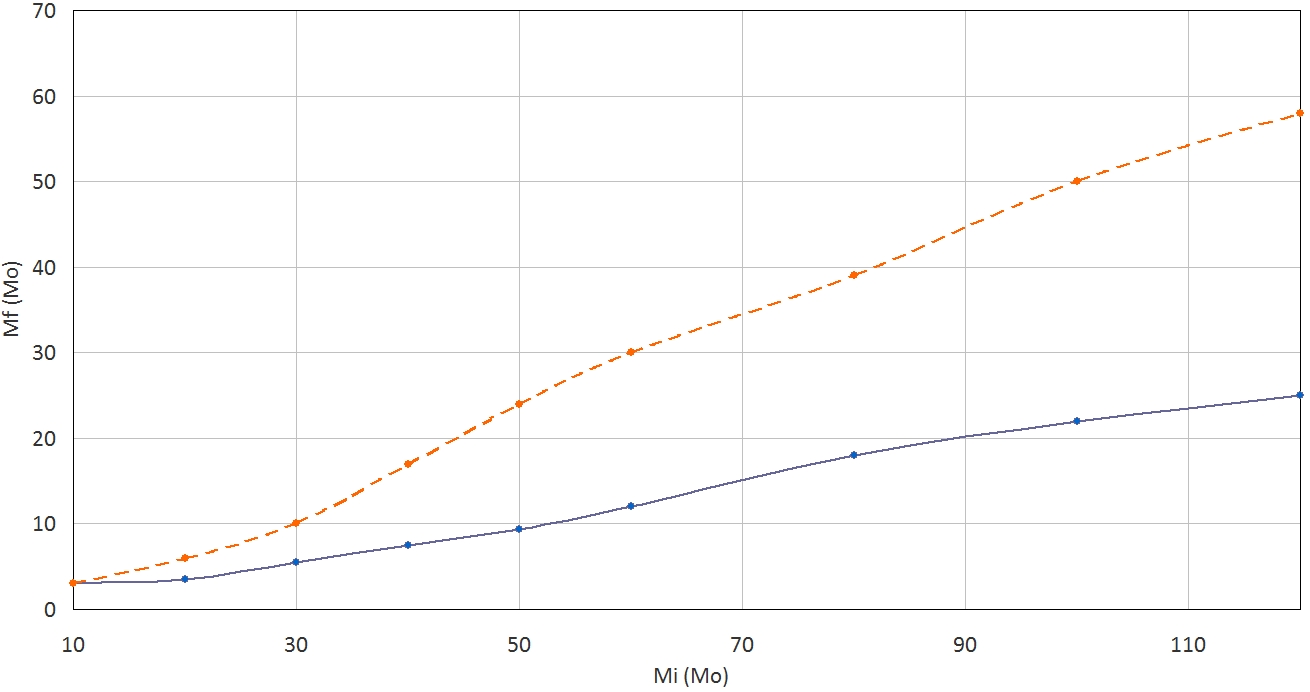}
     \caption{The relation between the initial mass of a massive primary in an interacting binary and its mass at the end of core helium burning (just prior to the formation of a NS or a BH). Solid blue: Solar. Dashed red: SMC. Masses in M$_{\odot}$.}
     \label{fig:5}
\end{figure}

\end{document}